\documentclass[letterpaper]{article}
\usepackage{aaai}
\usepackage{times}
\usepackage{helvet}
\usepackage{courier}
\usepackage{graphicx}

\long\def\remove#1{}
\long\def\comment#1{}

\newcommand{\figref}[1]{Figure~\ref{#1}}

\title{Information Contagion: an Empirical Study of the Spread of News on Digg and Twitter Social Networks}
\author{Kristina Lerman \and Rumi Ghosh  \\
USC Information Sciences Institute \\
Marina del Rey, CA 90292, USA }

\begin{document}

%       \email{lerman@isi.edu}
%       \email{rumig@usc.edu}
%       \email{khanduri@usc.edu}

\maketitle
\begin{abstract}
Social networks have emerged as a critical factor in information dissemination, search, marketing, expertise and influence discovery, and potentially an important tool for mobilizing people. Social media has made social networks ubiquitous, and also given researchers access to massive quantities of data for empirical analysis.  These data sets offer a rich source of evidence for studying dynamics of individual and group behavior, the structure of networks and global patterns of the flow of information on them. However, in most previous studies, the structure of the underlying networks was not directly visible but had to be inferred from the flow of information from one individual to another. As a result, we do not yet understand dynamics of information spread on networks or how the structure of the network affects it. We address this gap by analyzing data from two popular social news sites.  Specifically, we extract social networks of active users on Digg and Twitter, and track how interest in news stories spreads among them. We show that social networks play a crucial role in the spread of information on these sites, and that network structure affects dynamics of information flow.
\end{abstract}

%\keywords{Information Propagation, Social Dynamics, Web2.0, Digg, Twitter}

 % Claim: Empirical study of spread of news on Digg, Twitter, and the role social networks play in it.

\section{Introduction}
Social scientists have long recognized the importance of social networks in the spread of information~\cite{Granovetter}
and innovation~\cite{Rogers03}.
Modern communications technologies, notably email and more recently social media, have only enhanced the role of networks in marketing~\cite{Domingos01,Kempe03}, information dissemination~\cite{Wu03,Gruhl04}, search~\cite{Adamic05search}, and expertise discovery~\cite{Davitz07}. The recent DARPA Network Challenge\footnote{https://networkchallenge.darpa.mil} successfully tested the ability of online social networks to mobilize massive ad-hoc teams to solve real-world problems, which could potentially improve disaster response and coordination of relief efforts. In addition to making social networks ubiquitous, social media sites have given researchers access to massive quantities of data for empirical analysis.  These data sets offer a rich source of evidence for studying the structure of social networks~\cite{Leskovec08} and the dynamics of individual~\cite{Barabasi06b} and group behavior~\cite{Hogg09icwsm}, efficacy of viral product recommendation~\cite{Leskovec06}, global properties of the spread of email messages~\cite{Wu03,Liben-Nowell08pnas} and blog posts~\cite{Leskovec07blogs}, and identification of influential blogs~\cite{Gruhl04,Leskovec07kdd}. In most of these studies, however, the structure of the underlying network was not visible but had to be inferred from the flow of information  from one individual to another. This posed a serious challenge to our efforts to understand how the structure of the network affects dynamics of information spread on it.

Understanding this question is especially critical for the effective use of social media and peer production systems, which often aggregate over activities of, or contributions made by, many people in order to identify trending topics and noteworthy contributions.  Most of these sites also highlight activities of a person's social network links. Since people create links to others who are similar to them, or whose contributions they find interesting, the dynamics of information on a social network may be different from its dynamics within the general population. Separating in-network from out-of-network activity allows us, among other things, to better estimate the inherent quality of the contributions~\cite{CraneSornette08} or predict their future activity~\cite{Hogg10icwsm,Lerman08wosn}. This will in turn allow us to separate high quality contributions from noise.
% model of social dynamics

Social news sites Digg and Twitter offer a unique opportunity to study dynamics of information spread on social networks. Both sites have become important sources of timely information for people. The social news aggregator Digg allows users to \emph{submit} links to news stories and \emph{vote} on stories submitted by other users. On the microblogging service Twitter users \emph{tweet} short text messages that often contain links to news stories and comment on or \emph{retweet} messages of others. Both sites enable users to
%create social networks by explicitly declaring what other users they want to follow.
explicitly create links to other users they want to follow. Another important common feature is data transparency, with both sites providing programmatic access to detailed data about story and user activity.

This paper presents an empirical study of the role of social networks in the spread of information on Digg and Twitter.
For our study we  collected data about popular stories on Digg and Twitter that includes information about who voted or retweeted the story and when. In addition, we extracted the social networks of active users on these sites.
These data sets allow us to empirically characterize individual dynamics, network structure, and to map the spread of interest in news stories through the network.
First, we empirically characterize the structure of social networks on both sites. While the number of fans a user has on each site exhibits a long-tail distribution, Digg's social network is denser and more interconnected than Twitter's, as judged by the number of reciprocated links and the network clustering coefficient. We also show that user activity on both sites has a power-law distribution, albeit with different exponents. Next, we study evolution of the number of votes stories receive. We show that user interface affects dynamics of votes, with evolution of Digg stories going through two distinct stages. Nevertheless, the number of votes accumulated by stories on both sites saturates after a period of about a day to a value that reflects their popularity. Next, we study how information spreads through the social network by measuring how the number of in-network votes a story receives, i.e., votes from fans of the submitter or previous voters, changes in time. We show that the structure of the network affects dynamics of information spread, with information reaching nodes faster in a denser network of Digg than Twitter. However, Twitter stories spread farther, as judged by the total number of in-network votes they receive. We conclude with a discussion of implications of the study.

%The paper is organized as follows. First, we present an overview of Digg and Twitter, describe their interfaces and functionality. We then quantitatively characterize individual user activity and dynamics of story voting. Finally, we empirically characterize the spread of news on social networks and discuss implications.

\section{Social News}
\label{sec:social_news}
Social media has become an important channel for people to share information. On Digg, Twitter, Slashdot, Reddit, and Facebook, among others, users post news or links to news stories, discuss them, and share their opinions in real time. Often, these sites are the first to break important news.
%, and in some countries they are at times the only source of news.
%For example, the news of Rumsfeld's resignation in the wake of the 2006 U.S. Congressional elections broke on Digg at least 20 minutes before it was related by other news organizations~\cite{Rose2006Rumsfeld}.
After the Christmas 2009 failed attempt to blow up a US commercial airliner, Twitter was the first source to report new security measures for international flights~\cite{nytimes}. In addition to news, these sites are being used as a tool to organize  people. For example, in the aftermath of the disputed elections in Iran in June 2009, the opposition movement used Twitter to mobilize the public, organize protests, and inform people about the latest developments, which was more vital in the absence of reliable official information sources.

%Several sites, such as Digg, Slashdot and Reddit, then aggregate opinions of many users to select what they deem to be the most interesting or important stories. In addition, social media sites often enable users to discover news through their social networks, by seeing what stories their designated friends discovered recently. We extract social networks from two such sites, Digg and Twitter, and study their role in the dynamics of news propagation through the user community.

\textbf{Digg} (http://digg.com) is a popular social news aggregator with over 3 million registered users.  Digg allows users to submit links to and rate news stories by voting on, or \emph{digging}, them. There are many new submissions every minute, over 16,000 a day.  Digg picks about a hundred stories daily
%that it deems to be most \emph{popular}
to feature on its front page. Although the exact promotion mechanism is kept secret, it appears to take into account the number and the rate at which story receives votes. Digg's success is largely fueled by the emergent front page, created by the collective decisions of its many users.

A newly submitted story goes to the \emph{upcoming} stories list, where it remains for 24 hours, or until it is promoted to the \emph{front page}, whichever comes first. Newly submitted stories are displayed as a chronologically ordered list, with the most recent story at the top of the list, 15 stories to a page.
%To see older stories, a user must navigate to the upcoming stories page 2, 3, etc.
Promoted (or `popular') stories are also displayed in a reverse chronological order on the front pages, 15 stories to a page, with the most recently promoted story at the top of the list.
%The vast majority of users who visit Digg peruse the front page only.
The importance of being promoted has, among other things, spawned a black market\footnote{As an example, see http://subvertandprofit.com} which claims the ability to manipulate the voting process.
%To see older stories, user must navigate to front page 2, 3, etc. Figure~\ref{fig:screenshot} shows a screenshot of a Digg front page.

Digg also allows users to designate friends and track their activities. The \emph{friends interface} allows users to see the stories friends recently submitted or voted for. The friendship relationship is asymmetric. When user $A$ lists user $B$ as a \emph{friend}, $A$ can watch the activities of $B$ but not vice versa. We call $A$ the \emph{fan} of $B$. A newly submitted story is visible in the upcoming stories list, as well as to submitter's fans through the friends interface. With each vote it also becomes visible to voter's fans. The friends interface can be accessed by clicking on \emph{Friends Activity} tab at the top of any Digg page. In addition, a story submitted or voted on by user's friends receives a green ribbon on the story's Digg badge, raising its visibility to fans.

% data retrieval
We used Digg API to collect data about 3,553 stories promoted to the front page in June 2009. The data associated with each story contained story title, story id, link,  submitter's name, submission time, list of voters and the time of each vote, the time the story was promoted to the front page.
% the number of times it was viewed
%and the number of votes, or diggs, it received.
In addition, we collected the list of voters' friends. From this information, we were able to reconstruct the fan network of Digg users who were active during the sample period.

\textbf{Twitter} (http://twitter.com) is a popular social networking site that allows registered users to post and read short (at most 140 characters) text messages, which may contain URLs to online content, usually shortened by a URL shortening service such as bit.ly or tinyurl. A user can also retweet or comment on another user's post, usually prepending it with a string ``RT @\emph{x},'' where $x$ is a user's name. Posting a link on Twitter is analogous to submitting a new story on Digg, and retweeting the post is analogous to voting for it. Like Digg, Twitter allows users to designate as friends other users whose posts they want to follow. Being a \emph{follower} on Twitter is equivalent to being a fan on Digg.

% data retrieval
%Twitter limits access to user profile and search APIs to white-listed IP addresses only.
Twitter restricts large-scale access to its data to a limited number of entities. One of these, Tweetmeme (http://tweetmeme.com), aggregates all Twitter posts to determine frequently retweeted URLs, categorizes the stories these URLs point to, and presents them as news stories in a fashion similar to Digg's front page. We collected data from Tweetmeme using specialized page scrapers developed using Fetch Technologies's AgentBuilder tool. For each story, we retrieved the name of the user who posted the link to it, the time it was posted, the number of times the link was retweeted, and details of up to 1000 of the most recent retweets. For each retweet, we extracted the name of the user, the text and time stamp of the retweet. We were limited to 1000 most recent retweets by the structure of Tweetmeme. We extracted 398 stories from Tweetmeme that were originally posted between June 11, 2009 and July 3, 2009. Of these, 329 stories had fewer than 1000 retweets. Next, we used Twitter API to download profile information for each user in the data set. The profile included the complete list of user's friends and followers.

%Active users == users who have at least one vote (Digg) or one tweet (Twitter) during the time data was collected.
\begin{figure*}[bth]
  \begin{tabular}{cc}
  \includegraphics[width=3.1in]{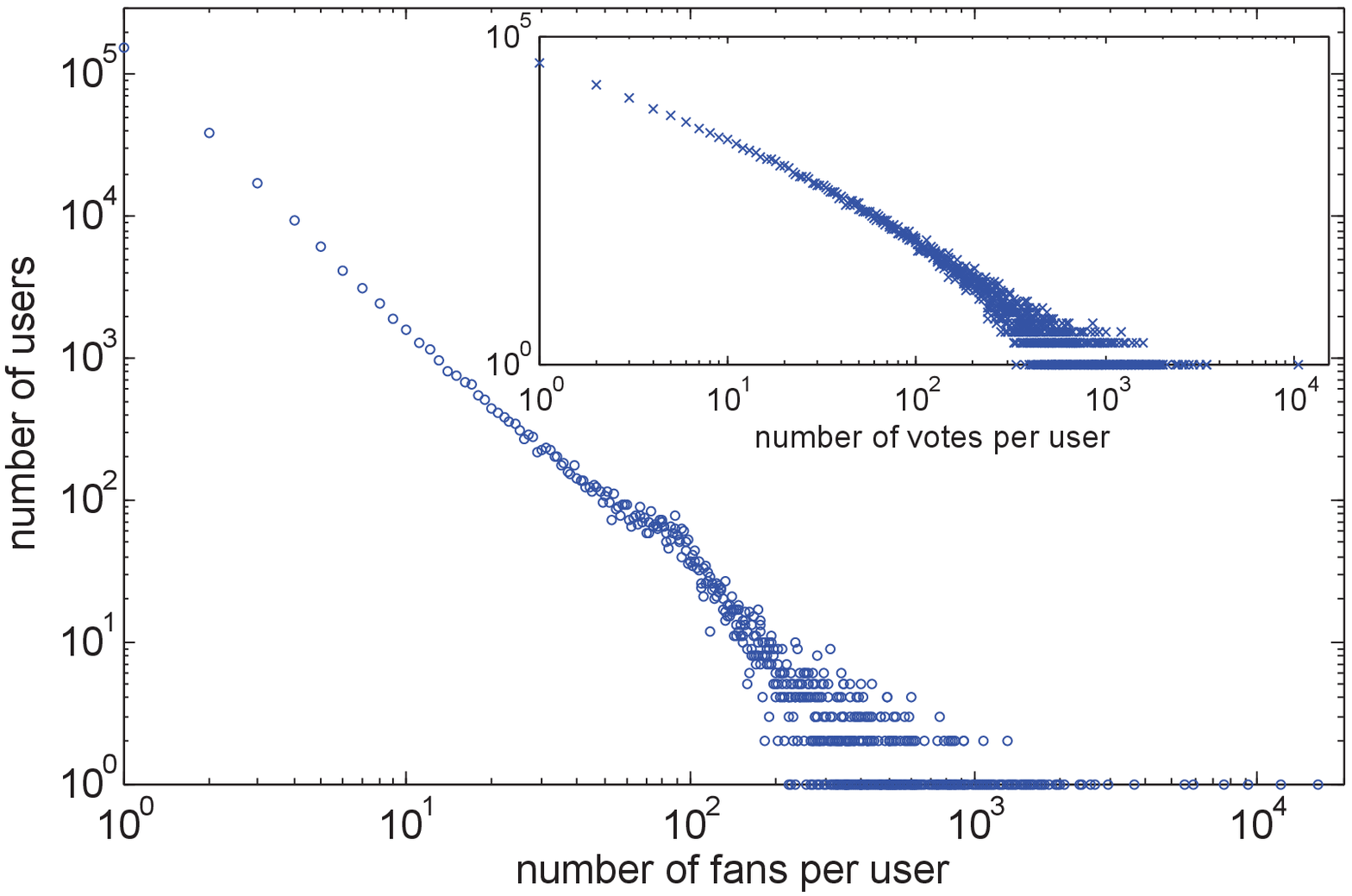}  &

  \includegraphics[width=3.1in]{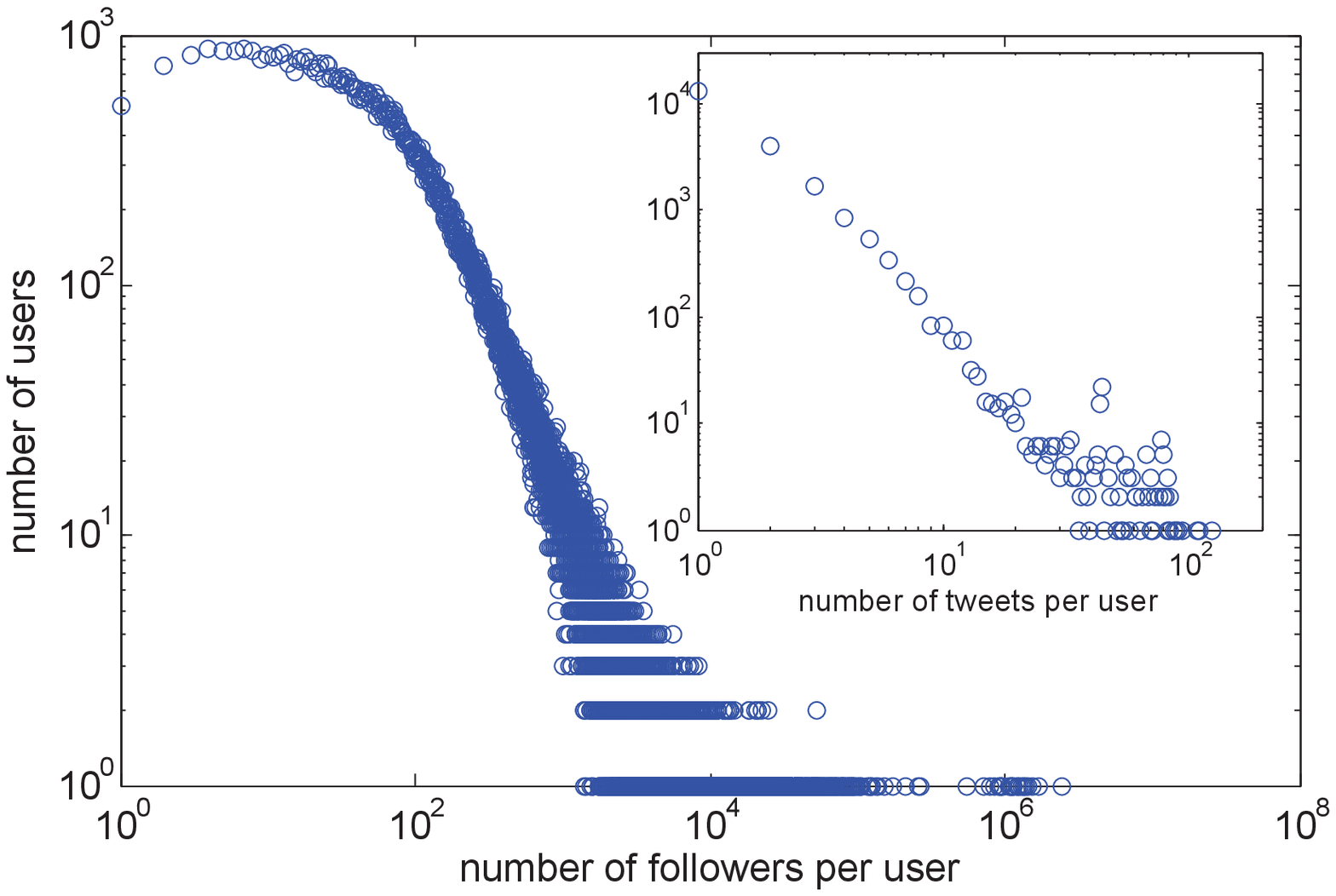}\\
    (a) Digg &(b) Twitter
  \end{tabular}
  \caption{Distribution of user activity. (a) Number of active fans per user in the Digg data set vs the number of users with that many fans. Inset shows distribution of voting activity, i.e., number of votes per user vs number of users who cast that many votes. (b) Number of active followers per user in the Twitter data set vs the number of users with that many followers. Inset shows distribution of retweeting activity.}\label{fig:fans}
\end{figure*}

\begin{figure*}[tbhp]
  \begin{tabular}{cc}
  \includegraphics[width=3.1in]{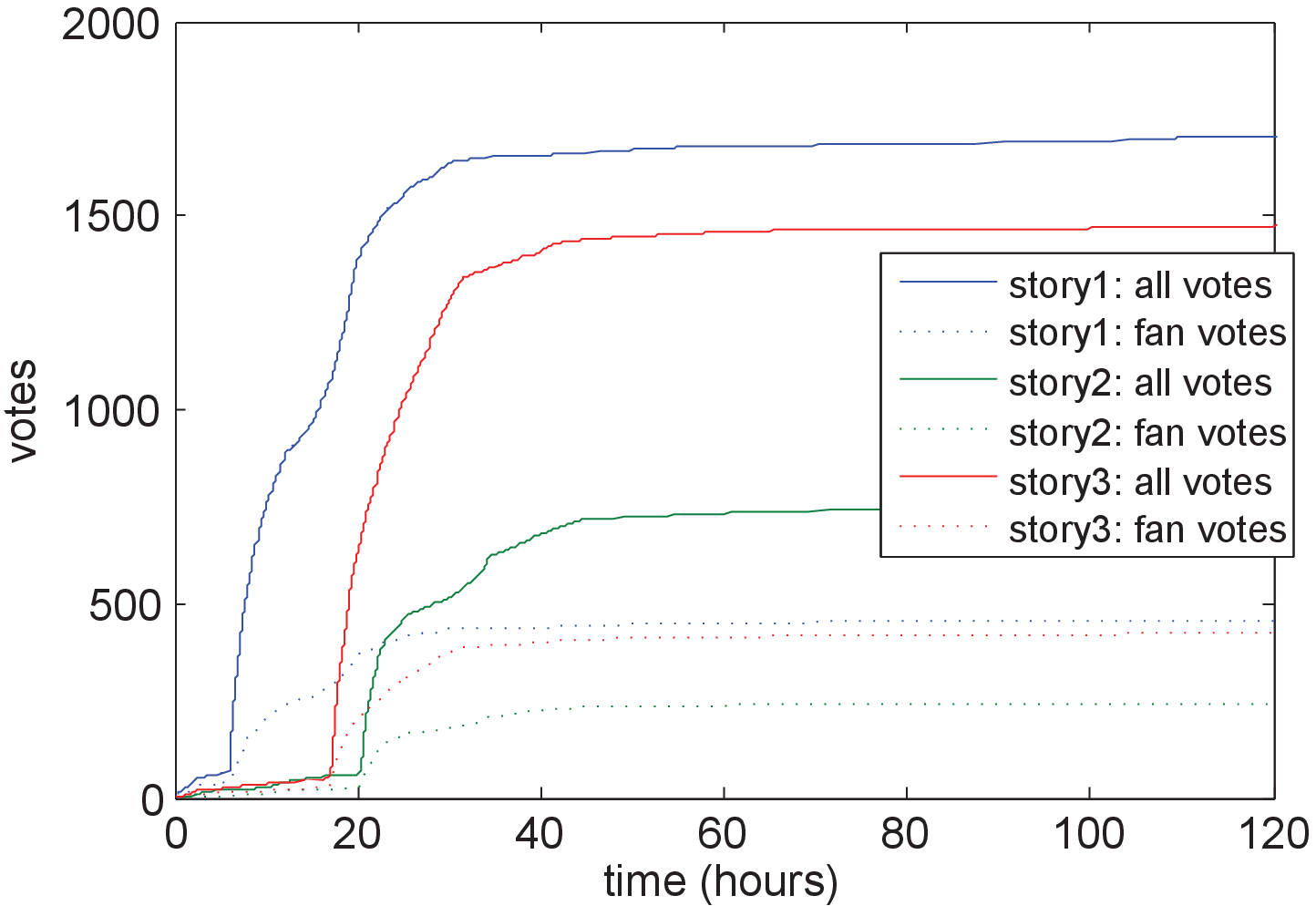}&
    \includegraphics[width=3.1in]{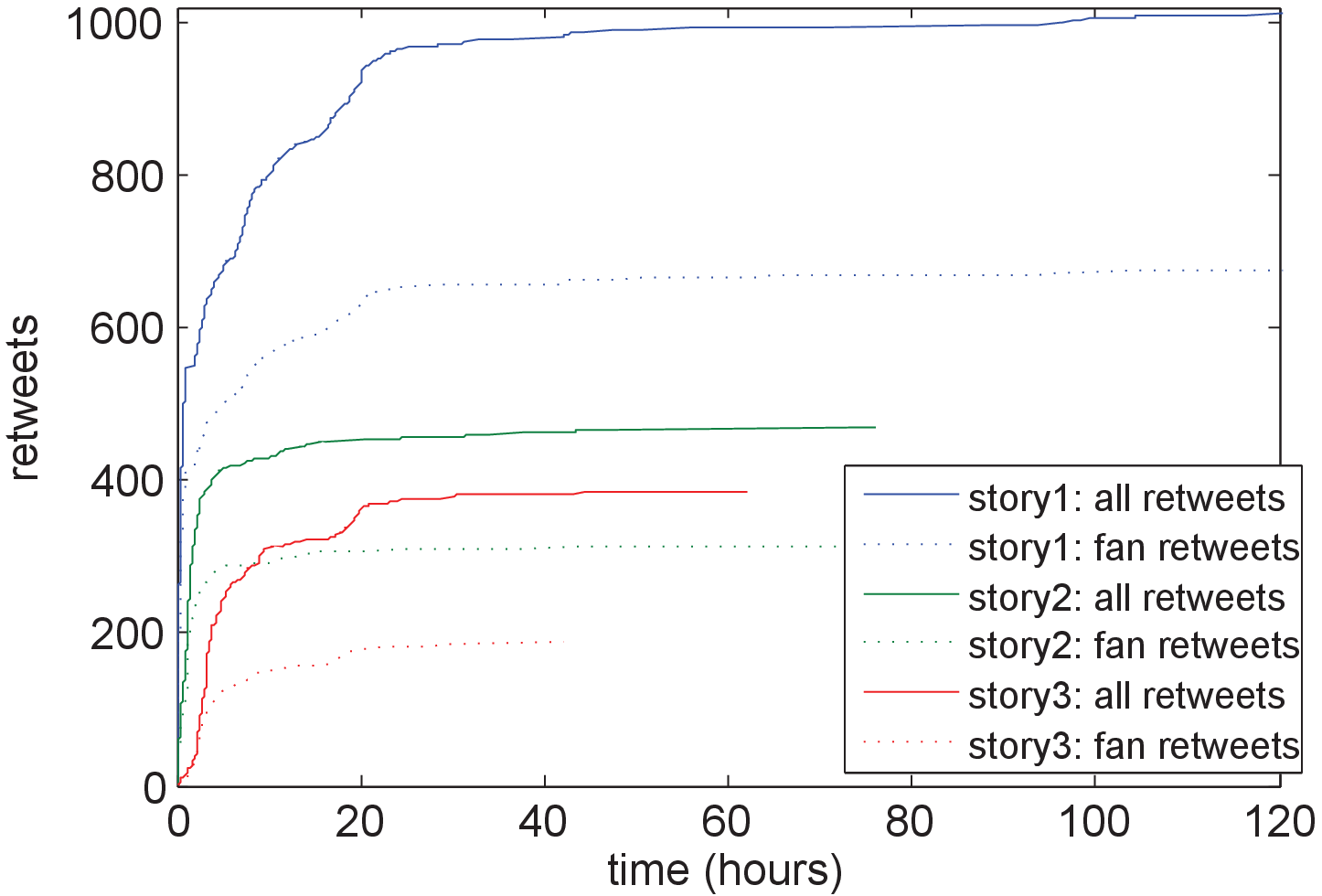}\\
   (a) Digg &
    (b) Twitter
  \end{tabular}
  \caption{Dynamics of stories on Digg and Twitter. (a) Total number of votes (diggs) and fan votes received by stories on Digg since submission. (b) Total number of times a story was retweeted and the number of retweets from followers since the first post vs time. The titles of stories on Digg were: story1: ``U.S. Government Asks Twitter to Stay Up for \#IranElection'', story2: ``Western Corporations Helped Censor Iranian Internet'', story3: ``Iranian clerics defy ayatollah, join protests.'' The titles of retweeted stories were: story1:``US gov asks twitter to stay up'', story2:``Iran Has Built a Censorship Monster with help of west tech'', story3:``Clerics join Iran's anti-government protests - CNN.com.''
}\label{fig:dynamics}
\end{figure*}

\section{Characteristics of User Activity}
\label{sec:user_activity}

%258,220 friends in friends table
% max 996 friends; min 1
% max fans: inactive(16216), kevinrose(12038), ofa(9291)
We define as \emph{active user} any user who voted for at least one story on Digg or retweeted at least one story on Twitter. There are 139,409 active Digg and 137,582 active Twitter users in our sample. On Digg, 71,834 active users designated at least one other user as a friend, with a total of 258,220 friend links. Active users on Twitter were connected to 6,200,051 users. From this data, we were able to reconstruct the fan networks of active users, i.e., active users who are watching activities of other users.
Figure~\ref{fig:fans} shows the distribution of number of active fans and followers per user. Digg's distribution, shown in Fig.~\ref{fig:fans}(a), has a long-tail shape that is common to degree distributions in real-world complex networks~\cite{Clauset09}.
%In other words, on Digg, few users have thousands of active fans, while many users have just a few, or one, fan.
%Users with the most active fans are \emph{kevinrose}, one of Digg founders, with 12,038 active fans, and \emph{ofa} with 9,291 active fans.
%This means that 12,038 and 9,291 of the 139,409 voters in our sample marked kevinrose and ofa respectively as a friend.
%We had 398 stories from 11th June 2009 to 3rd July 2009. Of these, 329 stories have votes less than 1000.
%Since Tweetmeme allows us to extract the maximum of 1000 votes per story, to study the complete dynamics of a story with time, we use these 329 stories.
%The database comprises of 137,582 distinct voters and  they are connected to 6,200,051 people. There are about  80,476,253 connections in the dataset.
%In the Twitter data set, there was a similar number of distinct voters, 137,582, but they were connected to a total of 6,200,051 people. Note that we were able to extract all followers from an active user's profile. Figure~\ref{fig:fans}(b) shows the distribution of number of followers per user.
Twitter's distribution, shown in Fig.~\ref{fig:fans}(b), has a peak at around 100 followers and a long tail.

% network structure
As the numbers above suggest, the Digg social network is denser, more tightly knit than the Twitter social network. We measure density by the number of reciprocal friendship links and the modified clustering coefficient. A reciprocal, or mutual, friendship link exists when user $A$ marks $B$ as friend and vice versa. There were 125,219 such links among 279,725 distinct users in the Digg sample and 3,973,892 mutual links among 6,200,051 users in the Twitter sample. Normalizing these counts by the number of all possible mutual links in the network gives us the fraction of mutual links $f_m$. For Digg $f_m=3.20\times 10^{-6}$, and for Twitter $f_m=2.07\times 10^{-7}$, an order of magnitude smaller. The clustering coefficient $f_c$ measures the degree to which a node's network neighbors are interlinked. We define the clustering coefficient for directed networks such as those that exist on Digg and Twitter as the fraction of closed triangles that exist out of all possible sets of three nodes, or triples. For simplicity, we define a closed triangle as a cycle of length three that exists when $A$ lists $B$ as a friend, $B$ lists $C$ and $C$ lists $A$ as a friend. There were 166,239 such triangles in the Digg network, giving us the clustering coefficient $f_c=7.60 \times 10^{-12}$, and 4,566,952 triangles on Twitter, giving the clustering coefficient of $f_c=1.92 \times 10^{-14}$ that is two orders of magnitude smaller.  Due to the size of the networks, we implemented these metrics using Hadoop\footnote{http://hadoop.apache.org/}. We suspect that the differences in density of the two networks are due to their age, since Twitter is a more recent service than Digg. With time, we expect the Twitter network to grow denser~\cite{Leskovec05} and become as tightly knit as Digg.

%\subsection{Diversity of user activity}
%Distribution of user activity: Number of votes a user made vs number of users who made that many votes
%(see Tad's pdf, p1)
%Question: is diversity (distribution) of user activity the same on Digg and Twitter?
Next, we characterize users' voting activity. The 139,409 active users in the Digg data set cast 3,018,197 votes on 3,553 stories.  User activity is not uniform, as shown in inset Fig.~\ref{fig:fans}(a). While majority of users cast fewer than 10 votes, some users voted on thousands of stories over the sample time period. The distribution of the number of retweets per user in the Twitter data set has a similar shape, with the number of retweets per user ranging from 1 to about 100. The difference in slopes in these distribution is likely explained by the level of effort~\cite{Wilkinson08} required to vote on Digg vs retweet on Twitter.

%\subsection{User activity and social connectedness}
%Are these correlated? Are more connected users more active?
%Next, we look at correlation between user's activity levels and connectedness: i.e., do more connected users tend to be more active? Comparing correlation between connectedness and activity of 50,133 Digg users: correlation between number of active fans and friends was $0.3445$, between active fans and votes was $0.3794$, and between active friends and votes $0.1420$. All correlations were statistically significant.
% Correlation between all fans and votes was $0.4506$ for 580 users we had data about.

\section{Dynamics of Voting}
\label{sec:dynamics_voting}
Our data sets contain a complete record of voting on Digg front page stories and frequently retweeted stories on Twitter.
%Each story's record contains the time it was submitted or first posted, as well as the names of voters and the time the vote (or retweet) was made, and for Digg stories, the time it was promoted to the front page.
From this data we can reconstruct dynamics of voting. In addition to voting history, we also know the active fan network of Digg and Twitter users and use this information to check whether a particular voter is a fan of the submitter or previous voters. We call these in-network votes \emph{fan votes}. This information allows us to study how interest in the story spreads through the social networks on Digg and Twitter.

%\subsection{Votes vs time}
%Number of total votes a story receives vs time
Figure~\ref{fig:dynamics}(a) shows the evolution of the number of votes received by three Digg stories about post-election unrest in Iran in June 2009. While the details of the dynamics differ, the general features of votes evolution are shared by all Digg stories and can be described by a stochastic model of social voting~\cite{Hogg09icwsm}. While in the upcoming stories queue, a story accumulates votes at some slow rate. The point where the slope abruptly changes corresponds to promotion to the front page. After promotion the story is visible to a large number of people, and the number of votes grows at a faster rate. As the story ages, accumulation of new votes slows down~\cite{Wu07} and finally saturates.
%The total number of votes the story receives by that time gives a measure of its success or {popularity}.
Figure~\ref{fig:dynamics}(b) shows the evolution of the number of times stories on the same topics were retweeted.
%Since there is no ``upcoming'' vs ``promoted'' stage on Twitter,
The number of retweets grows smoothly until it saturates. It takes about a day for the number of votes/retweets to saturate on both sites.

\begin{figure*}[tbh]
\begin{center}
 \begin{tabular}{cc}
  \includegraphics[height=2.0in]{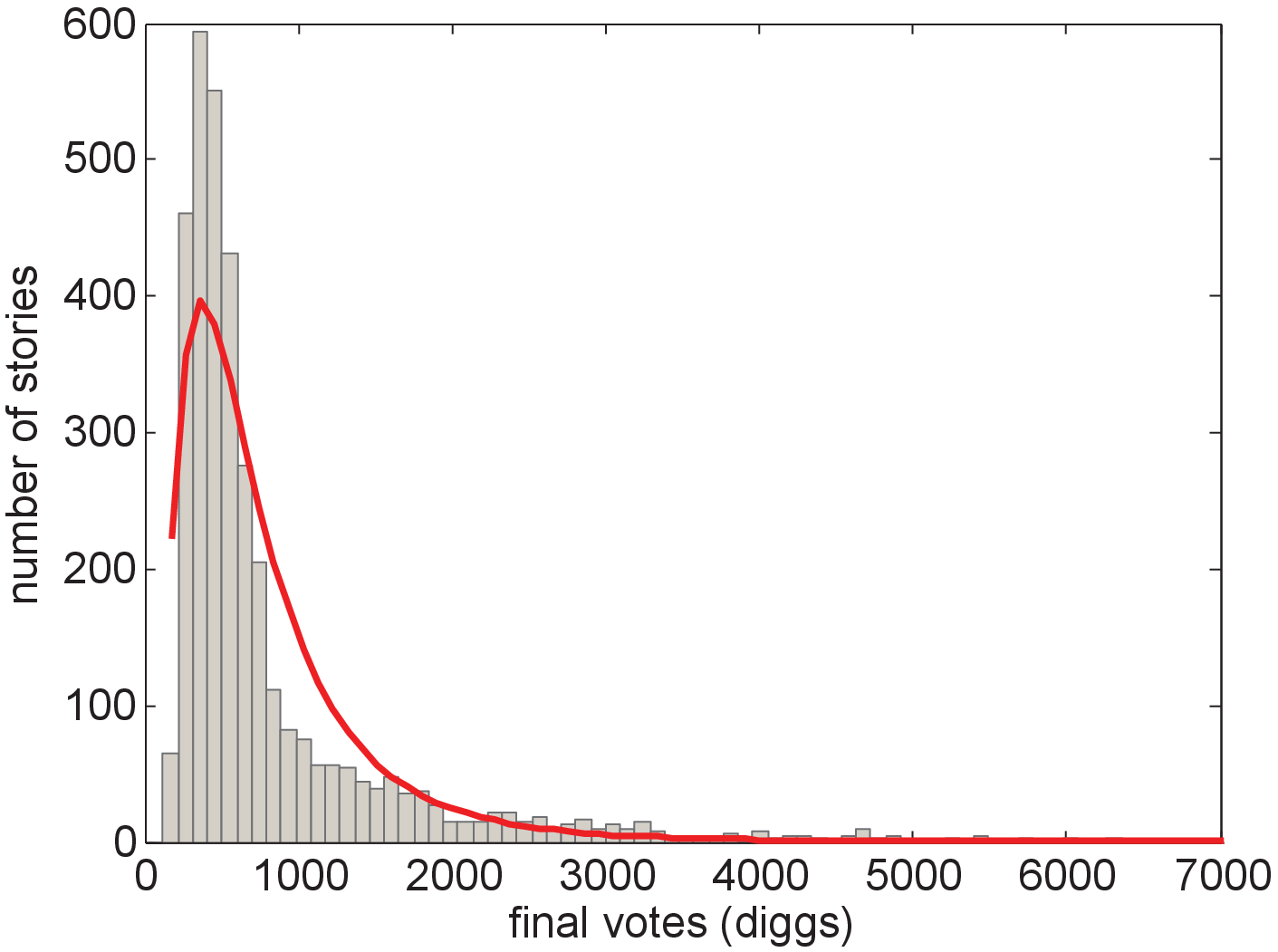}
  &
  \includegraphics[height=2.0in]{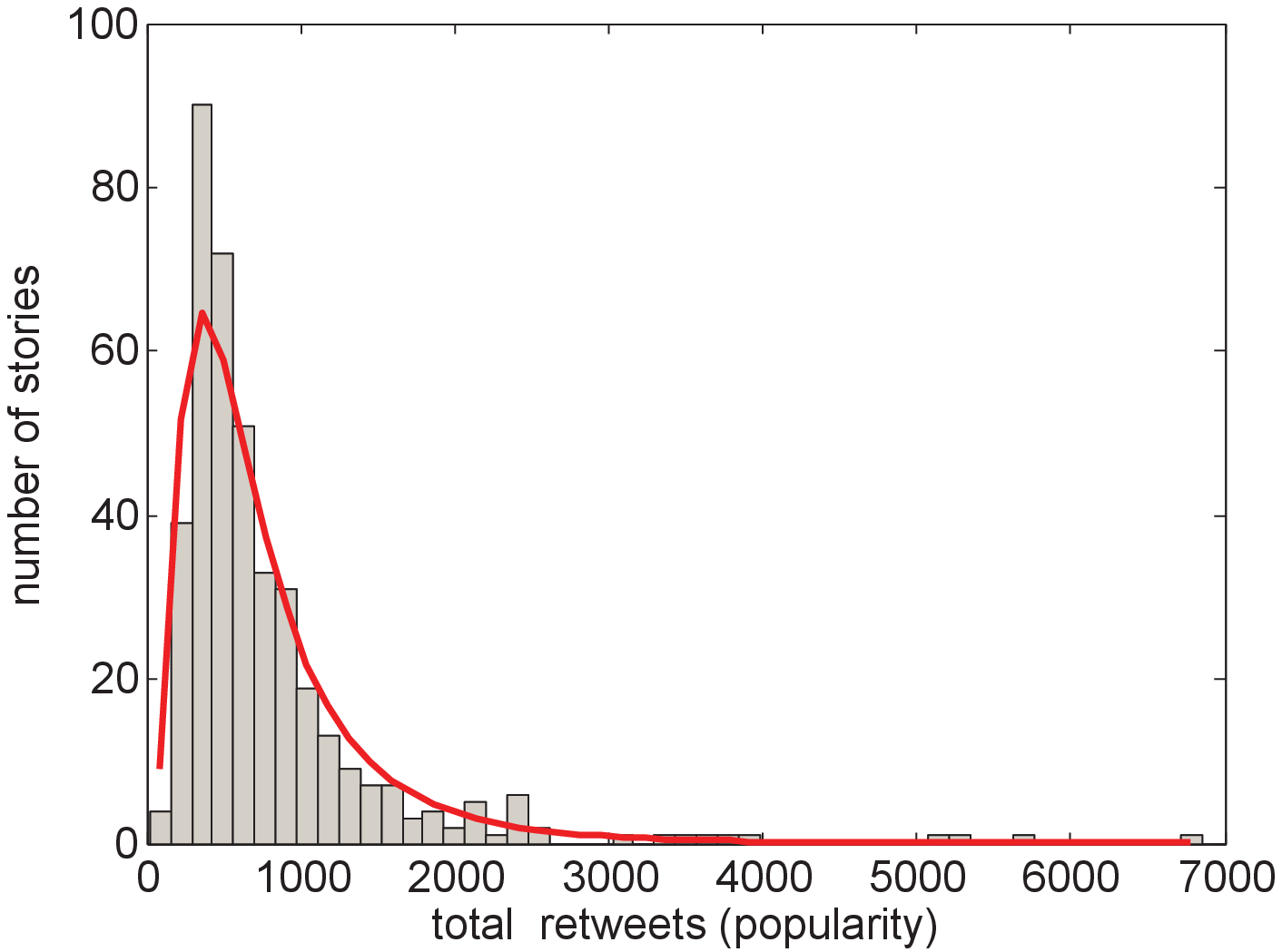} \\
  (a) Digg &
   (b) Twitter
  \end{tabular}
\end{center}
  \caption{Distribution of story popularity. (a) Distribution of the total number of votes received by Digg stories, with line showing log-normal fit. The plot excludes the 15 stories that received more than 6,000 votes. (b) Distribution of the total number of times stories in the Twitter data set were retweeted, with the line showing log-normal fit. }\label{fig:votes_histo}
\end{figure*}

\begin{figure*}[tbh]
\begin{center}
  \begin{tabular}{cc}
  \includegraphics[height=1.8in]{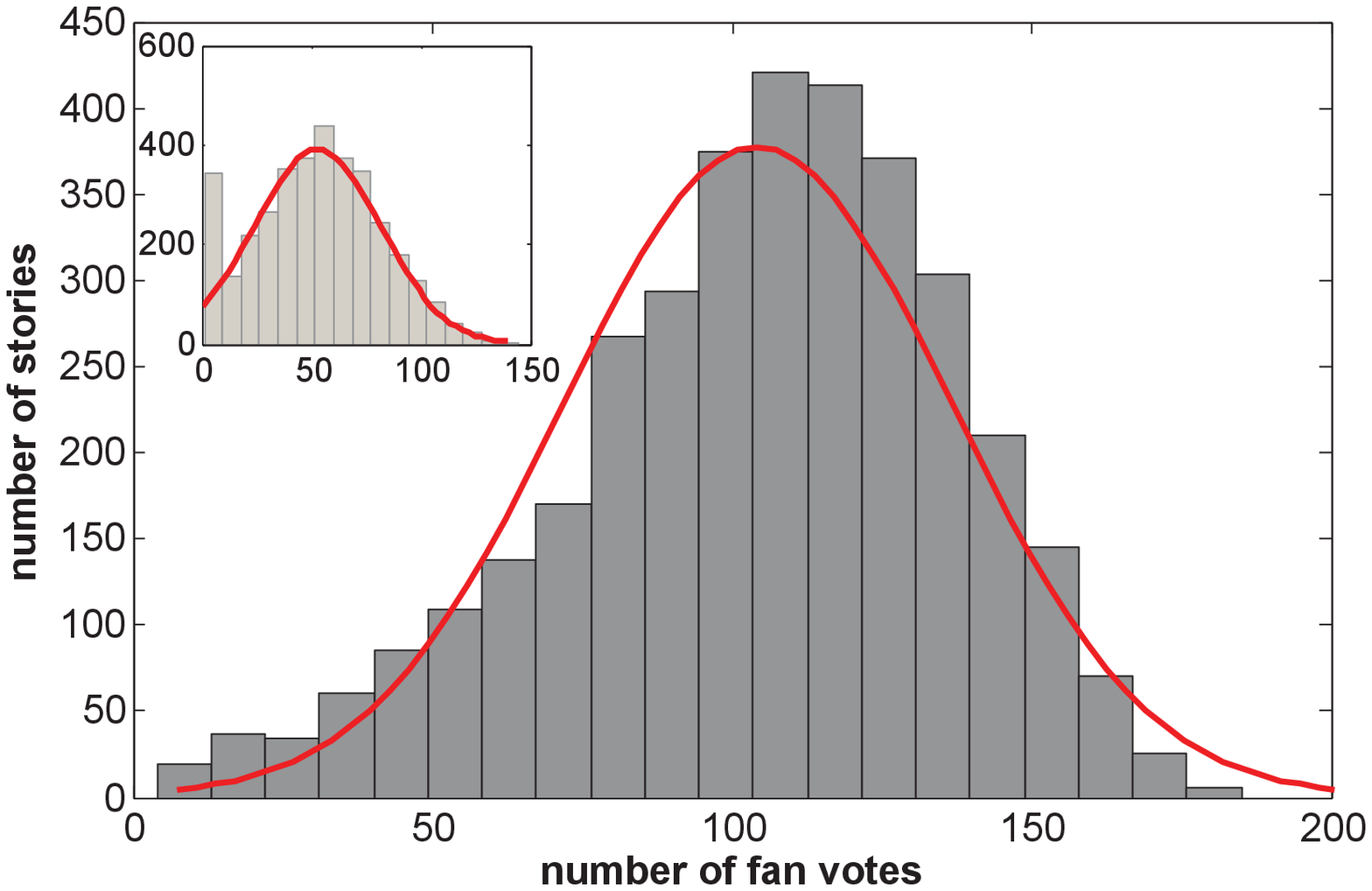}
  &
  \includegraphics[height=1.8in]{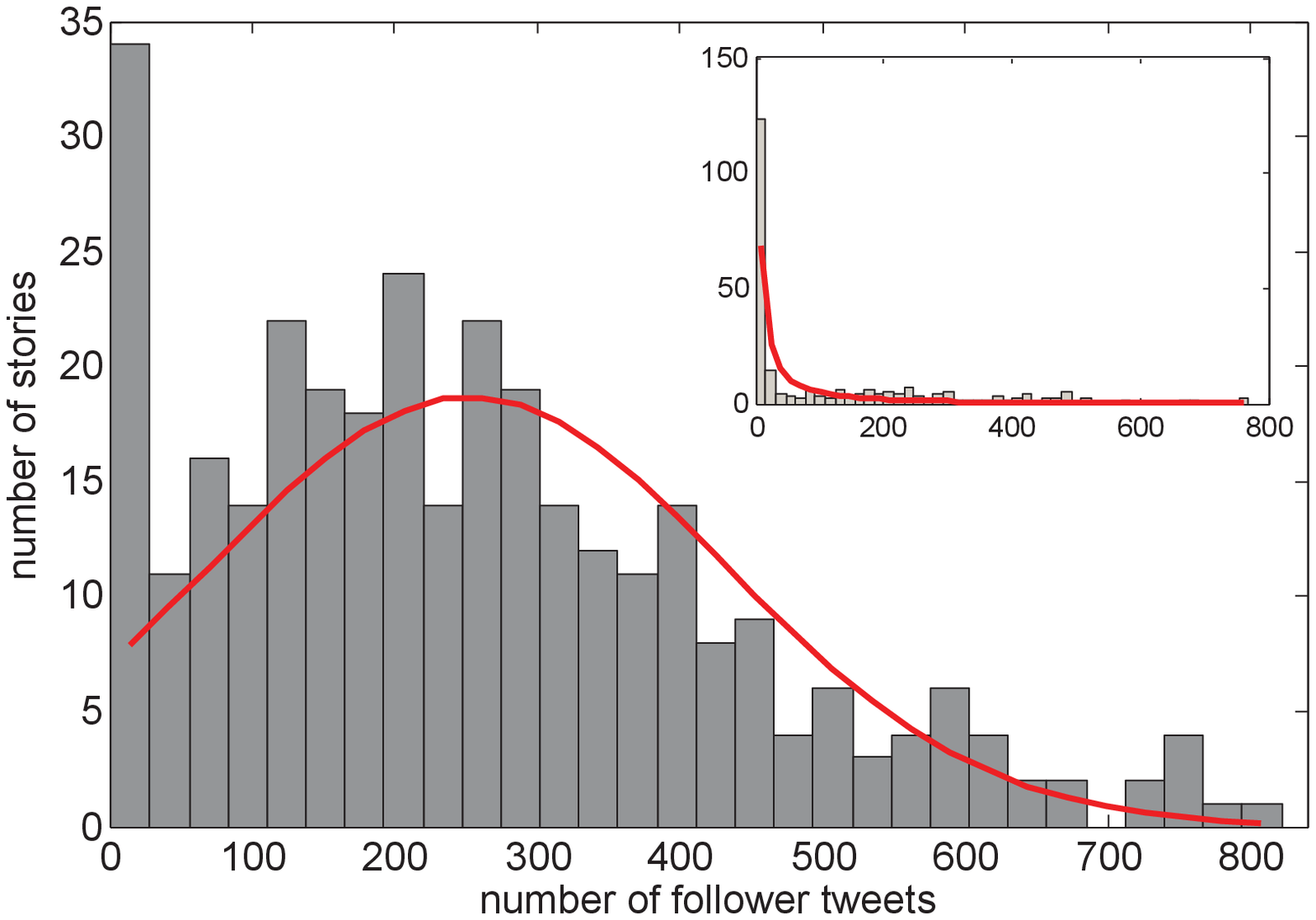} \\
  (a) Digg &
  (b) Twitter
  \end{tabular}
\end{center}
  \caption{Distribution of story cascade sizes. (a) Histogram of the distribution of the total number of fan votes received by Digg stories (size of the interest cascade). The inset shows the distribution of the number of votes from submitter's fans. (b) Histogram of the distribution of the total number of retweets from followers. The inset shows the distribution of the number of retweets of a story from submitter's followers. }\label{fig:fanvotes_histo}
\end{figure*}

\paragraph{Distribution of popularity}
%\label{sec:distribution_of_votes}
%Observations: severe inequality in popularity. Some stories become extremely popular, most only moderately so.
The total number of times the story was voted for and retweeted reflects their popularity among Digg and Twitter users respectively.
%Figure~\ref{fig:votes_histo} shows the distribution of popularity of stories on both sites.  The plot excludes the 15 stories that received more than 6,000 votes.
The distribution of story popularity on either site, Figure~\ref{fig:votes_histo}, shows the `inequality of popularity'~\cite{Salganik06}, with relatively few stories becoming very popular, accruing thousands of votes, while most are much less popular, receiving fewer than 500 votes.\footnote{This distribution applies to Digg's front page stories only. Stories that are never promoted to the front page receive very few votes, in many cases just a single vote from the submitter. } The most common number of votes by a story is around 500 on Digg and 400 on Twitter. These values are well described by a lognormal distribution (shown as the red line in the figure). %, with maximum likelihood estimates of the mean and standard deviation of $\log(votes)$ equal to $6.42$ and $0.72$ respectively.

%We discuss the implications of this in greater detail in Section~\ref{sec:quality}.

The log-normal distribution of story popularity is typical of the ``heavy-tailed'' distributions associated with social production and consumption of content. In a heavy-tailed distribution a small but non-vanishing number of items generate uncharacteristically large amount of activity. These distributions have been observed in a variety of contexts, %~\cite{anderson06},
including voting on Digg~\cite{Wu07} and Essembly~\cite{HoggSzabo09icwsm}, edits of Wikipedia articles~\cite{Wilkinson08}, and music downloads~\cite{Salganik06}.
%Log-normal distributions are a result of a multiplicative process~\cite{Mitzenmacher04}. In the case of Digg, a simple multiplicative process for generating votes~\cite{Hogg09icwsm} is the following: a user's choice to vote for a story depends on the story's \emph{visibility} and its \emph{interestingness}. {Visibility} is the probability the user will see the story during a visit, and it is heavily affected by social influence, i.e., through the information about choices of other users, which Digg highlights prominently both through the friends interface and by promoting the story to the front page.  Story interestingness is the probability the user will follow the link to a story she sees and vote for it. This probability depends on user's interest in the topic of the story and also story quality.
Understanding the origin of such distributions is the next challenge in modeling user activity on social media sites.

\section{Dynamics of Voting on Networks}
\label{sec:network-dynamics}

\remove{
\begin{figure*}[hp]
  \begin{center}
  \begin{tabular}{cc}
  \includegraphics[width=2.8in]{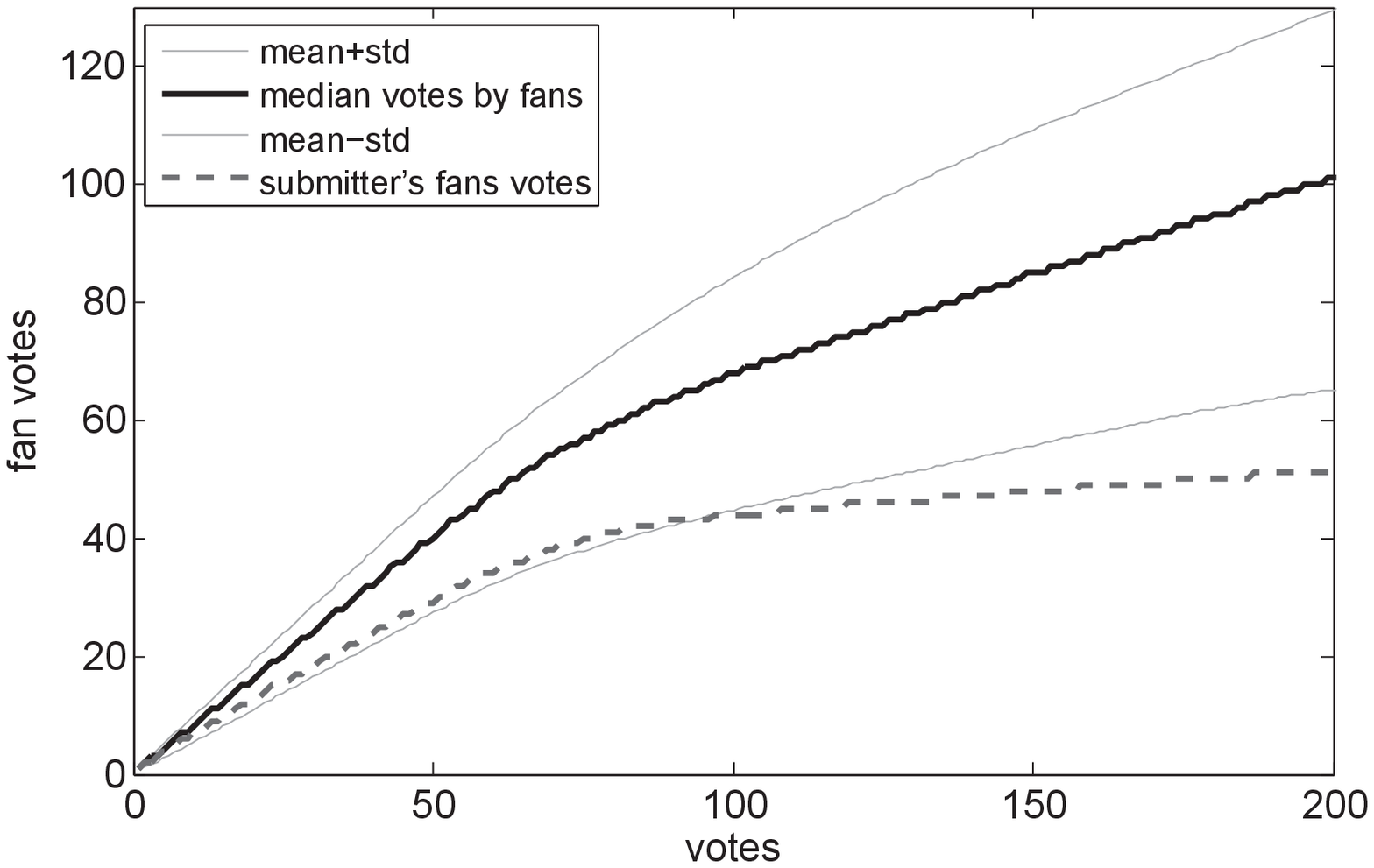}
  &
  \includegraphics[width=2.8in]{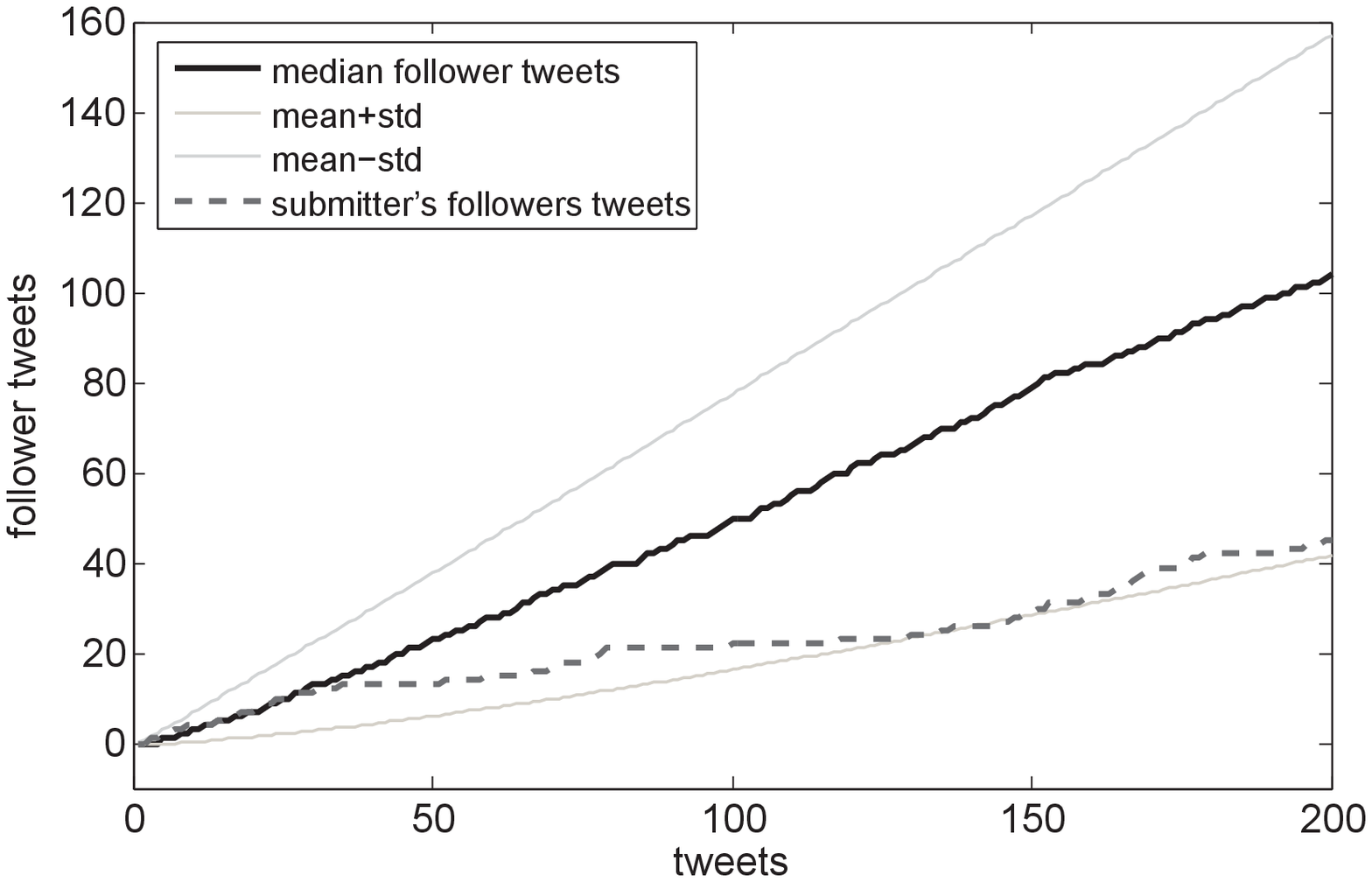}
  \\
  (a) & (b) \\
  \includegraphics[width=2.8in]{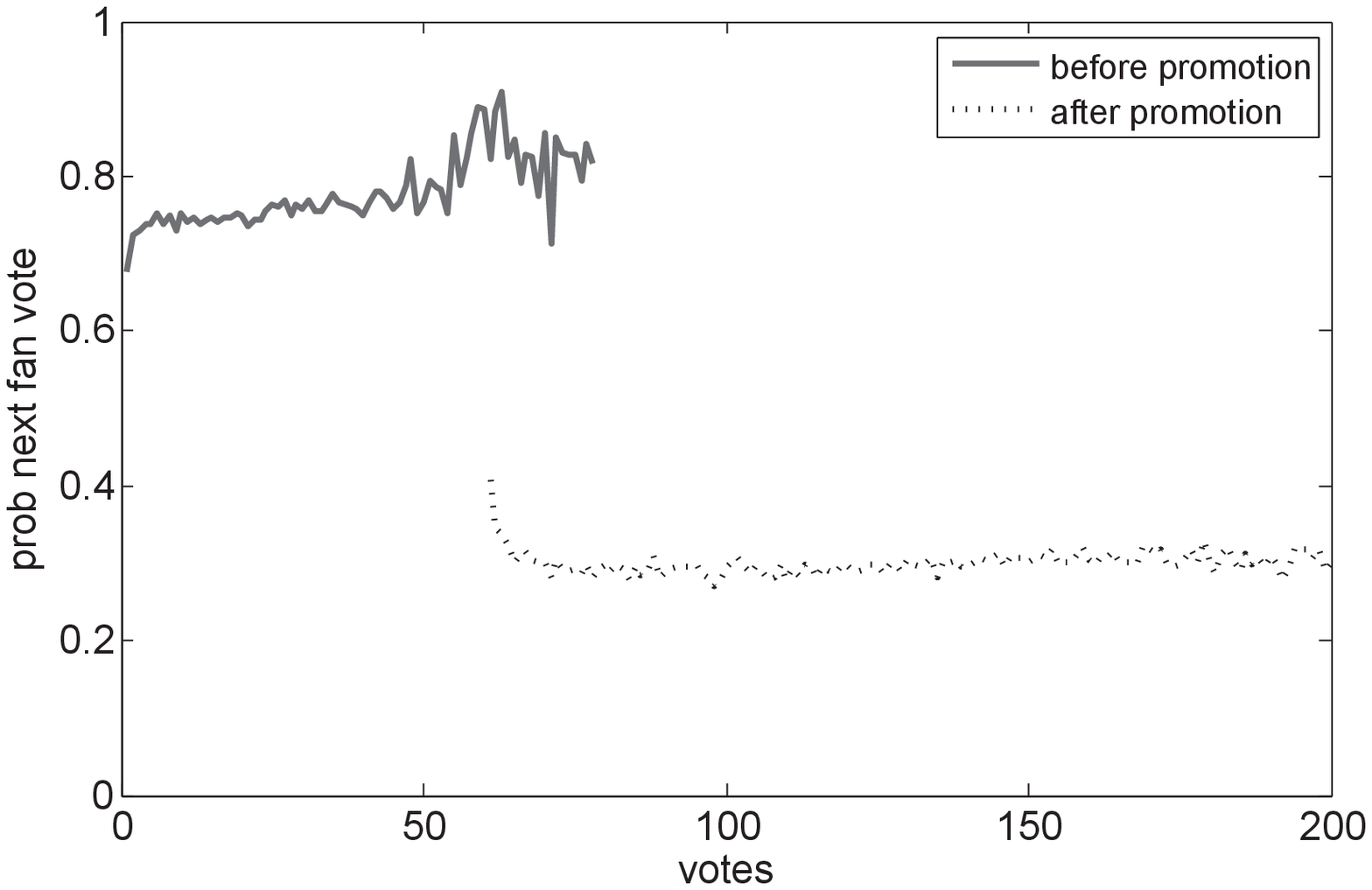}
  &
  \includegraphics[width=2.8in]{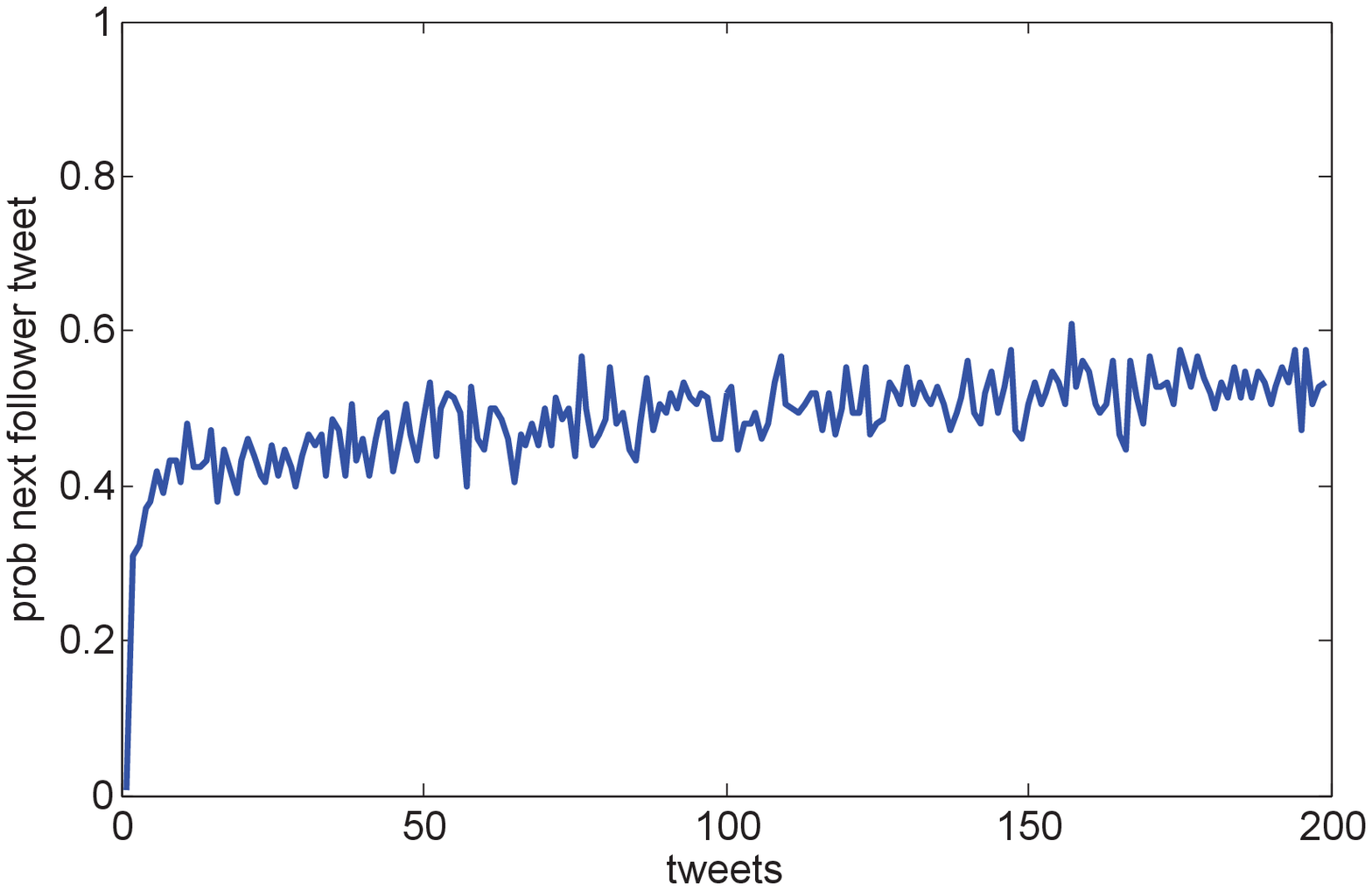}
  \\
  (c) Digg & (d) Twitter
  % twitter
  \end{tabular}
  \end{center}
  \caption{Spread of interest in stories through the network. (a) Median number of fan votes vs votes, aggregated over all Digg stories in our data set. Dotted lines show the boundary one standard deviation from the mean. Dashed lines shows the number of votes from fans of submitter. (d) Median number of retweets from followers vs all retweets, aggregated over all stories in the Twitter data set. (c) Probability next vote is from a fan before and after the story is promoted. (d) Probability next retweet is from a follower.}\label{fig:network_spread}
\end{figure*}
}

\remove{
%\subsubsection{Growth in fan visibility}
%This shows how new votes expose the story to new people via the social network.

%Plot number of fans vs votes. Likely need two plots for Distinct fans and all fans. p8 of Tad's pdf.

%Plot number of new fans per vote vs votes. Shows a power-law decay in Tad's plot (p8)
}

At the time of submission, a Digg story is visible on the upcoming stories list and to submitter's fans through the friends interface. As users vote on the story, it becomes visible to their own fans via the friends interface.  Analogous to the spread of a contagious disease~\cite{newman02}, interest in the story cascades through the social network. When the story is promoted to the front page, it becomes visible to many nonfans, although users are still able to pick out stories their friends liked through the green ribbon on the story's Digg badge. Similarly, a new post on Twitter is visible to submitter's followers, and every user who retweets the story broadcasts it to her own followers. Although aggregators like Tweetmeme attempt to identify popular stories on Twitter in Digg-like fashion, there is no evidence that they boost their visibility to nonfans.

We can trace the cascade of interest in a story through the underlying social network of Digg (Twitter)
by checking whether a new vote (retweet) came from a fan (follower) of any of the previous voters, including the submitter.
%Although we cannot confirm that the voter actually discovered the story through the social network, if a link exists between the new voter and any of the previous voters, we assume that the new voter became aware of the story through this link.
We call such votes or retweets \emph{fan votes}, regardless of whether we are talking about Digg or Twitter. Therefore, the cascade (``information contagion'' in the title of this article) starts with story's submitter and grows as the story accrues fan votes.
%Cascades allow us to track the flow of information on a network.
Researchers have studied information cascades in email chain letters~\cite{Wu03,Liben-Nowell08pnas} and blog posts~\cite{Gruhl04,Leskovec07blogs} in order to obtain insights into the structure of the network, identify influential nodes within it, or predict popularity of content~\cite{Lerman08wosn}. Characterizing information cascades is necessary for creating a model of the dynamics of information on networks.

\begin{figure*}[tb]
  \begin{tabular}{cc}
  \includegraphics[width=2.9in]{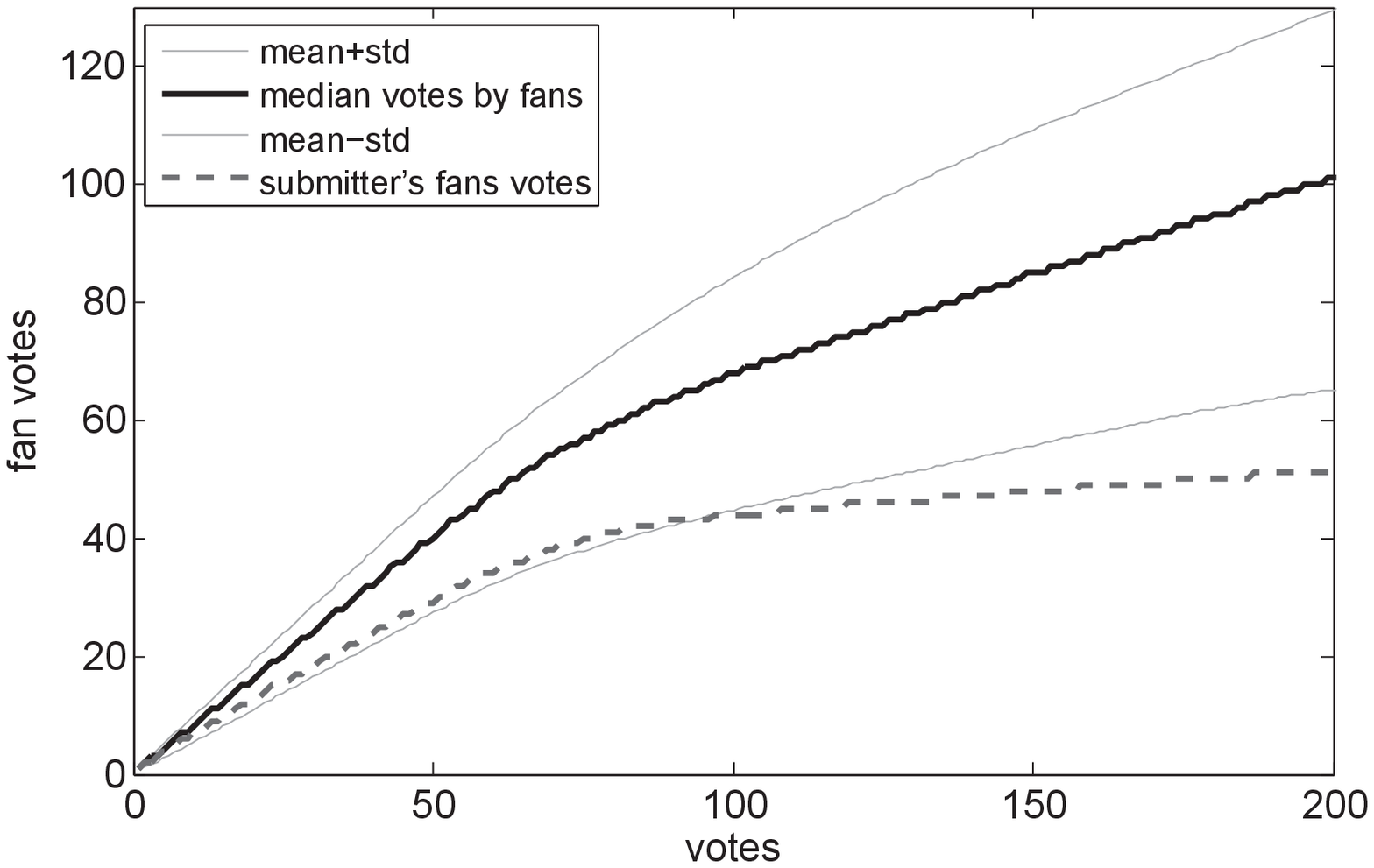} &
  \includegraphics[width=2.9in]{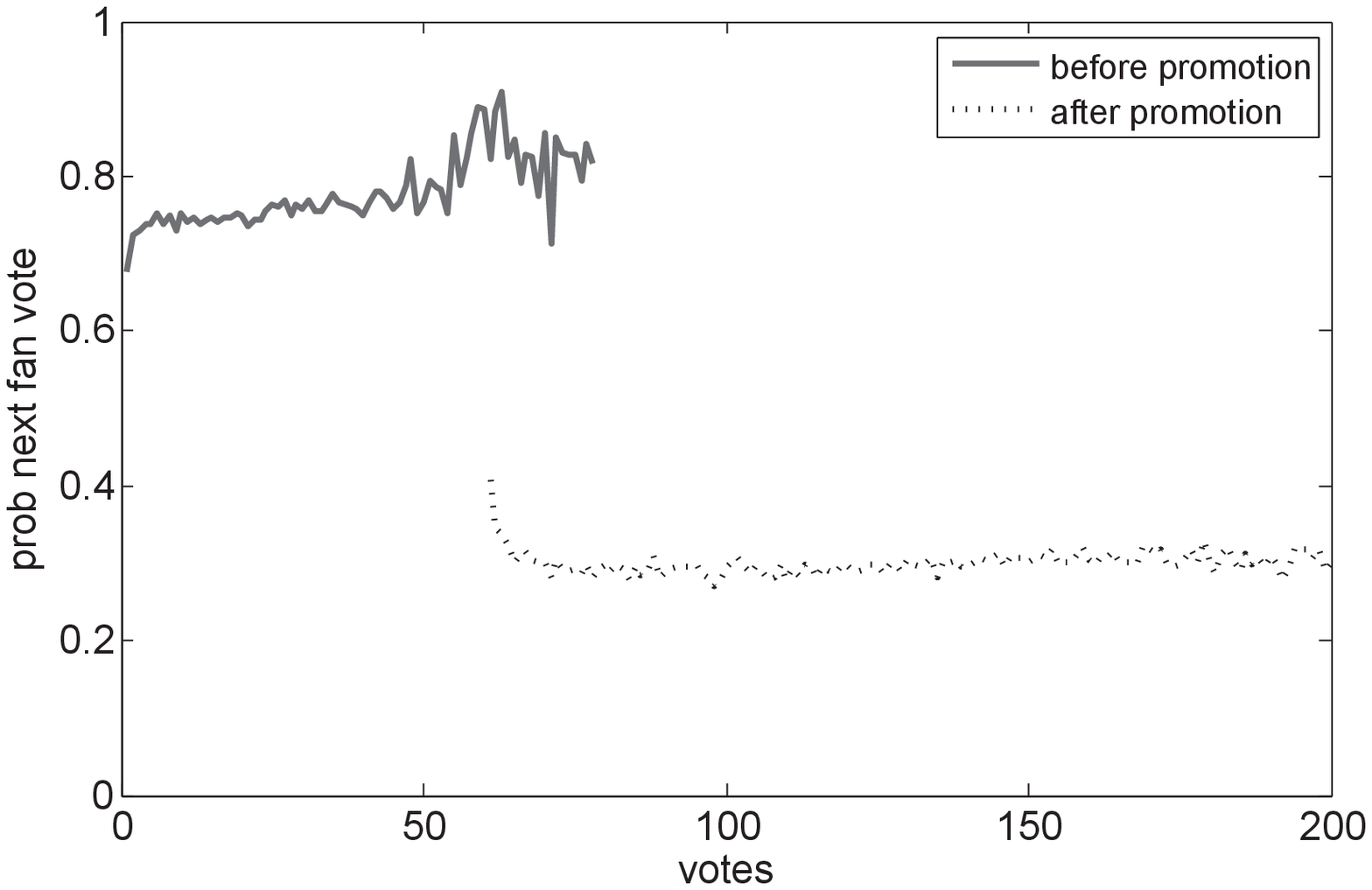}
  \\
  (a) &
  (b)
  \\
    \includegraphics[width=2.9in]{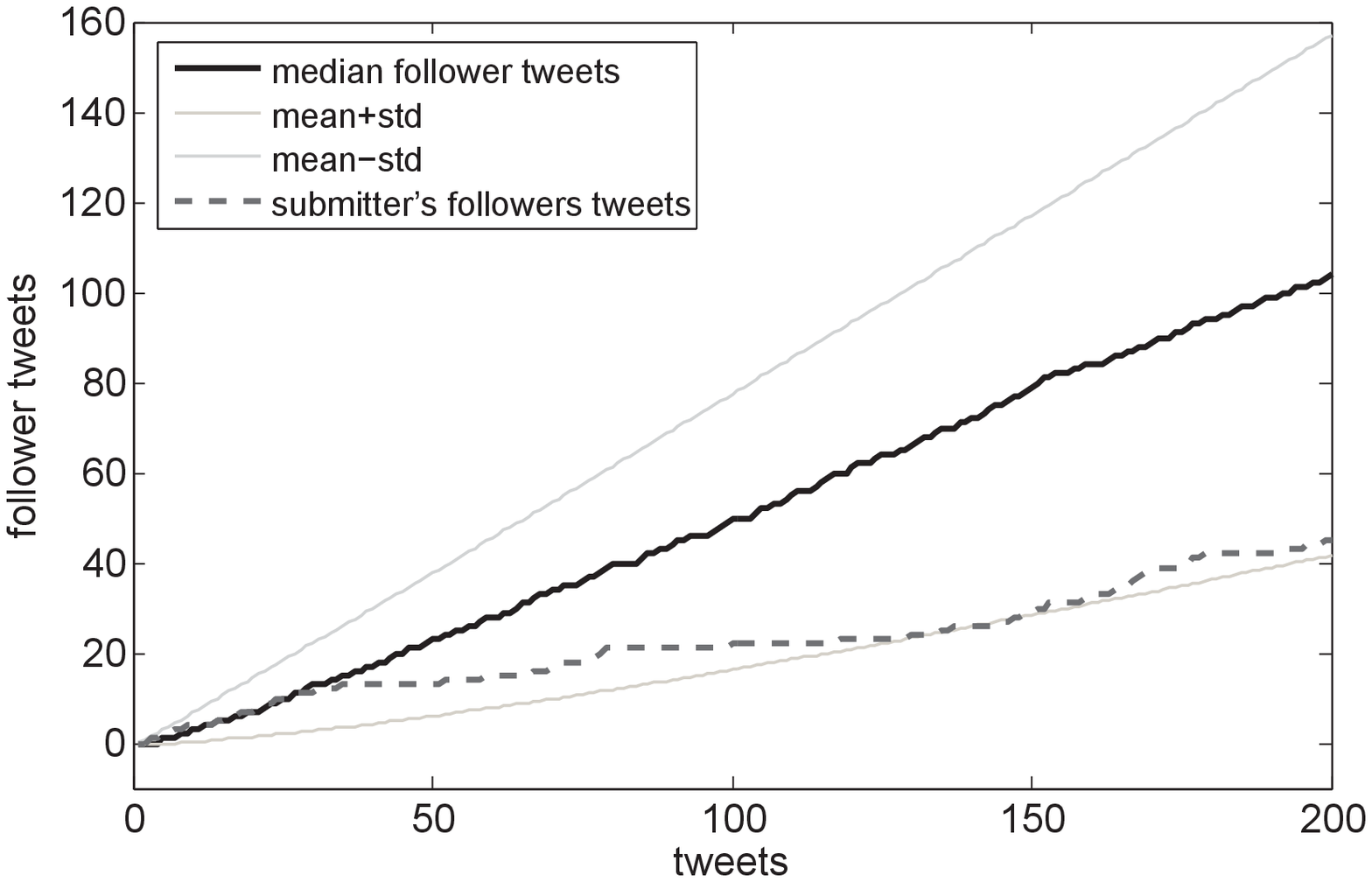} &
  \includegraphics[width=2.9in]{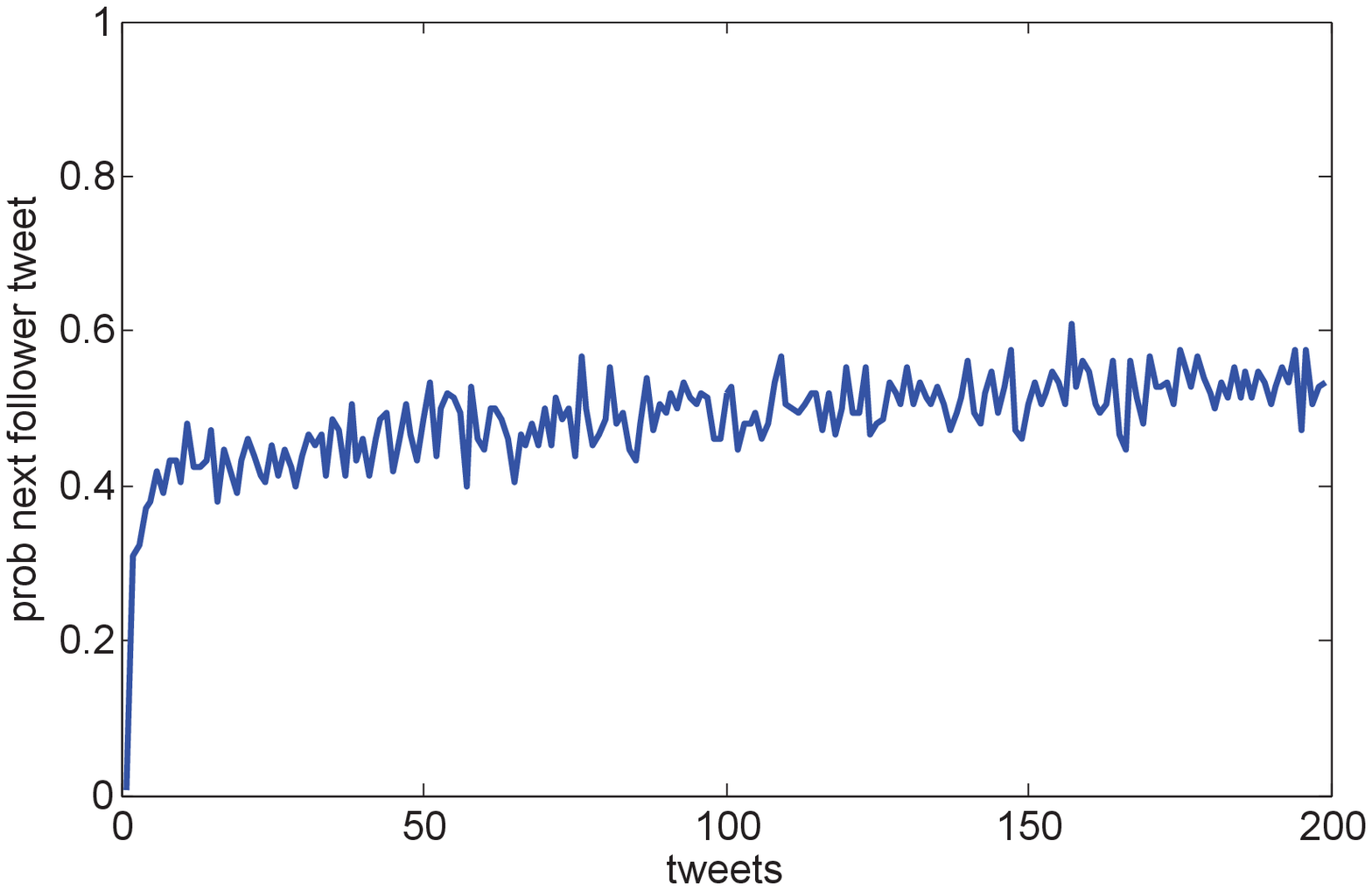}
  \\
  (c) &  (d)
  \end{tabular}
  \caption{Spread of interest in stories through the network. (a) Median number of fan votes vs votes, aggregated over all Digg stories in our data set. Dotted lines show the boundary one standard deviation from the mean. Dashed lines shows the number of votes from fans of submitter. (b) Probability next vote is from a fan before and after the Digg story is promoted. (c) Median number of retweets from followers vs all retweets, aggregated over all stories in the Twitter data set. (d) Probability next retweet is from a follower.}\label{fig:network_spread}
\end{figure*}

\remove{
\begin{figure*}[tbhp]
  \begin{tabular}{cc}
  \includegraphics[width=3in]{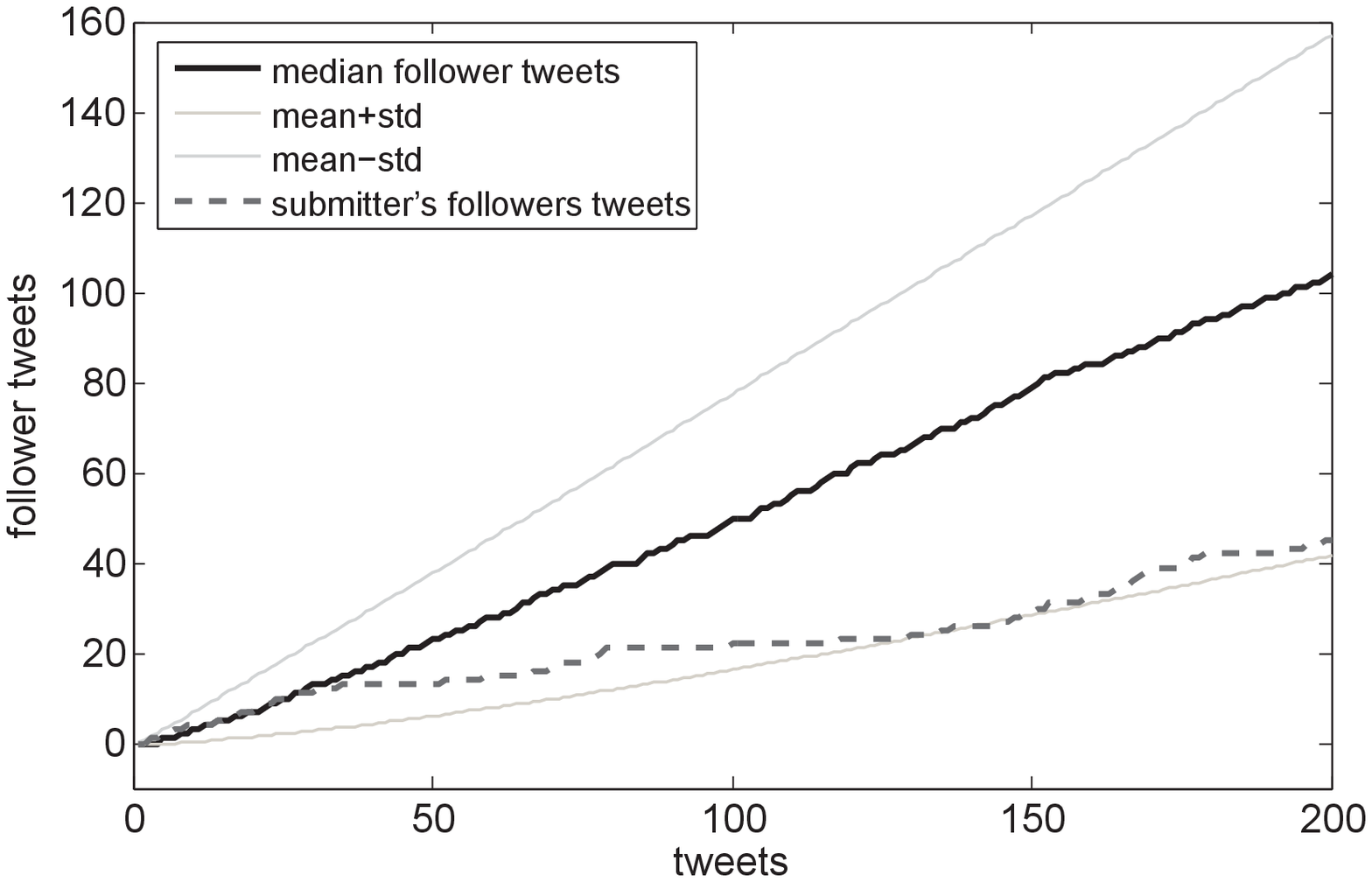} &
  \includegraphics[width=3in]{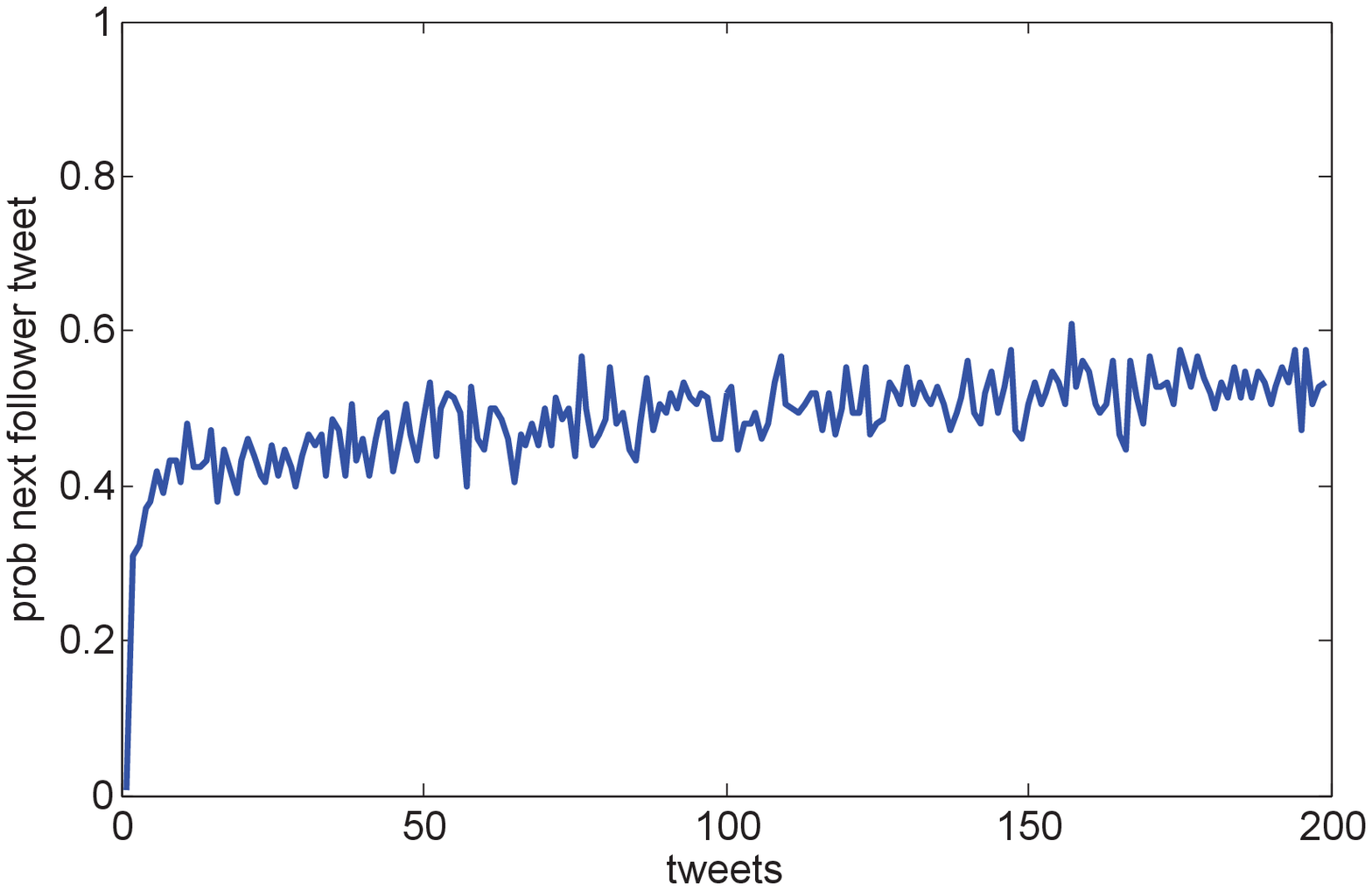}
  \\
  (a) &  (b)
  \end{tabular}
  \caption{Spread of interest in stories through the Twitter network. (a) Median number of tweets from followers vs all tweets, aggregated over all stories in the Twitter data set. (b) Probability next tweet is from a follower.}\label{fig:network_spread_twitter}
\end{figure*}
}

\paragraph{Dynamics and distribution of fan votes}
The dashed lines in Figure~\ref{fig:dynamics} show how the number of fan votes received by each story,
%the cascades, given by the number of fan votes,
grows in time.  Their evolution is similar to that of all votes and growth saturates after a period of about a day. The value at which growth saturates shows the story's range, or how widely it penetrates the social network.
%is the size of the cascade.
%In the story's early stages (before promotion for Digg stories), a higher proportion of votes are from fans than when the story is older.
%\subsection{Distribution of fan votes}
%\label{sec:distribution_fan_votes}
% KL - this is cascade size distribution. Leskovec07blogs gives power law distribution of blog post network
Figure~\ref{fig:fanvotes_histo} shows the distribution of cascade sizes generated by Digg and Twitter stories. These distributions are markedly different from the distribution of story popularity shown in Fig.~\ref{fig:votes_histo}. Although the distribution of network cascades of Digg stories, Fig.~\ref{fig:fanvotes_histo}(a), is slightly asymmetrical, it is best described by a normal with the mean and standard deviation equal to $104.27$ and $32.31$ votes respectively, not the log-normal distribution in Fig.~\ref{fig:votes_histo}(a).  It is also unlike distribution of cascade sizes in a blog post network, which has a  power law distribution~\cite{Leskovec07blogs}. Remarkably, there are no stories that did not generate a cascade, i.e., which did not receive any fan votes.

The inset in \figref{fig:fanvotes_histo}(a) shows the distribution of votes from submitter's fans only. It is also described by a normal function with a mean around 50 votes. A small fraction of stories, fewer than 400, did not have any votes from submitter's fans. This indicates that active users who are fans of the submitter are also fans of other voters, i.e., that the social network of active Digg users is dense and highly interlinked. This observation is supported by the finding of a relatively high clustering coefficient of the Digg social network.

The distribution of cascade sizes of of Twitter stories is shown in Fig.~\ref{fig:fanvotes_histo}(b). These also appear to be normally distributed, although a substantial number of stories do not spread on the network. This distribution is broader than that of Digg stories, which indicates that stories spread farther on the Twitter network. The distribution of the number of votes cast by submitter's followers, shown in inset in Fig.~\ref{fig:fanvotes_histo}(b), is markedly different from Digg. The vast majority of the stories did not receive any votes from submitter's followers, indicating that submitter's and other voters' followers are disjoint. This observation is supported by our finding that the Twitter social network is sparsely interconnected.
% must be a better way to say this

%Observations:
%\begin{itemize}
%  \item Votes saturate  - story popularity
%  \item Saturation within a day on Digg, Twitter sometimes more than a day
%  \item Digg has different visibility profile: stories visible on the upcoming stories pages (foot of the graph), then promoted to the front page (inflection point)
%\end{itemize}

\paragraph{Evolution of fan votes}
%Plot fan votes vs votes, averaged over all stories.
%Tad's plot on p5 for 2006 Digg data shows a good piecewise linear fit.

Figure~\ref{fig:network_spread} shows how the number of fan votes (size of the cascade),
aggregated over all stories, grows during the early stages of voting or retweeting. While there is significant variation in the number of fan votes received by a story,
%size of each cascade,
the aggregate exhibits a well-defined trend. The solid black lines show the median cascade size, while thin gray lines show the envelope of the boundary that is one standard deviation from the mean.

The
%number of fan votes
cascade
grows steadily with new votes on Digg (Fig.~\ref{fig:network_spread}(a)), although faster initially, indicating that there are two distinct mechanisms for story visibility on Digg.
%before promotion, which generally happens when a story accrues between 50 and 100 votes, than after promotion.
This is seen more clearly in Fig.~\ref{fig:network_spread}(b), which shows the probability that next vote is a fan vote and will increase the size of the cascade. We separate votes cast before promotion from those cast after the story is promoted. Before promotion, this probability is almost constant, at $p=0.74$. After promotion, it decays to a lower, but also almost constant value $p=0.3$.
This is consistent with our hypothesis that before promotion
%a  Digg story is mainly visible through the social network.
social networks are the primary mechanism for spreading interest in new stories.
Although a story is also visible on the upcoming stories list, few users actually discover stories there.
%, as new stories are quickly submerged by new submissions.
With 16,000 daily submissions, a new story is quickly submerged by new submissions and is pushed to page 15 of the upcoming stories list within the first 20 minutes. Few users are likely to navigate that far~\cite{huberman99}.
%a new story moves to page two of the upcoming stories list after two minutes, to page three after three minutes, etc. Few  few users are likely to navigate to page 15~\cite{huberman99} of the upcoming stories list to see a story submitted less than 20 minutes ago.
Promotion to the front page, which generally happens when a story accrues between 50 and 100 votes, exposes the story to a large and diverse audience, making social networks less of a factor in its spread, since large numbers of Digg users who read front page stories do not befriend others.
%As these users begin to vote, relative rate at which it accumulates new fan votes slows down.
%Once a story reaches the front page, social networks are no longer the primary mechanisms for spread interest in it.

%how the number of fan votes received by stories on Twitter changes as a function of all votes received, aggregated over all stories in the data set. The
The spread of interest in stories through the Twitter network, shown in Figure~\ref{fig:network_spread}(c), is similar to Digg. As on Digg, the median
number of fan votes
%received by stories in our data set cascade size
rises steadily during the early stages of voting. However, the rate of growth is nearly constant, indicating there is a single significant mechanism for making stories visible to voters, namely the social network. The probability that next retweet is from a fan, shown in Fig.~\ref{fig:network_spread}(d),  rises slowly from around $p=0.4$ to $p=0.55$. This value is lower than pre-promotion probability of next fan vote on Digg.
%, but lower than its post-promotion value.
The rate of interest spread appears to depend on the density of network. Initially, Digg stories spread faster through the social network than stories on Twitter, because of Digg's denser network structure, but after promotion they spread much slower as unconnected users see and vote on the stories.

%Looking at submitter's fans only, dashed line in Fig.~\ref{fig:network_spread_twitter}(a), shows that most of the initial fan votes come from submitter's fans. After this period submitter's fans do not vote as often, although half of the votes are coming from within social network.

% submitters fans
%As mentioned earlier, fan votes do not come from submitter's fans only.
The dashed lines in Fig.~\ref{fig:network_spread}(a) \& (c) show how the median number of votes from submitter's fans or followers changes with voting. By the time a story accumulates 50 votes on Digg (at which point some of the stories are promoted to the front page), about half of the votes are from submitter's fans, and another 10 are from fans of prior voters but not the submitter. After a story receives about 100 votes (by which point most of the stories are promoted),
the number of votes from submitter's fans changes very slowly, while the number of fan votes continues to grow. This indicates that submitter's fans vote for the story during its early stages and that users pay attention to the stories their friends submit. On Twitter, initial votes are from submitter's fans, but slows significantly later.

\section{Related Work}
\label{sec:related}
Several researchers studied dynamics of information flow on networks, however, empirical studies have produced conflicting results. \cite{Wu03} examined patterns of email forwarding within an organization and found that email forwarding chains terminate after an unexpectedly small number of steps. They argued that unlike the spread of a virus on a social network, which is expected to reach many individuals, the flow of information is slowed by decay of similarity among individuals within the social network. They measured similarity by distance in organizational hierarchy between the two individuals within an organization, or in general, as a number of edges separating two nodes within a graph.
Similarly, in a large-scale study of the effectiveness of word-of-mouth product recommendations, \cite{Leskovec06} found that most recommendation chains terminate after one or two steps. However, authors noted sensitivity of recommendation to price and category of product,  leaving open the question whether social networks are an effective tool for disseminating information, rather than purchasing products.
% On Digg, individuals are closer, but interest does not spread as far as on Twitter.
Contrary to these studies, we find that information, such as news, reaches many individuals within a social network. Moreover, the reach of information spread does not seem to depend on similarity between users, at least when similarity is measured by number of edges between them. On Digg, whose users are highly interconnected, a  story does not reach as many fans as on Twitter, where users are less densely connected.

Like Wu et al., \cite{Liben-Nowell08pnas} studied the patterns of forwarding of two popular email petitions. Unlike their expectations, the forwarding chains produced long narrow, rather than bushy wide, trees. In these studies, however, the structure of the underlying social network was not directly visible but had to be inferred by observing new signatures on the forwarded petitions. This method offers only a partial view of the network and does not identify all edges between individuals that participated in the email chain. If an individual has already forwarded the message, she will not do so again, and an edge between her and the sender will not be observed. In our study, on the other hand, the networks are extracted independently of data about the spread of information.

% blog space
A number of researchers have studied the flow of information and influence in the blogosphere and in a virtual world.
\cite{Gruhl04} traced topic propagation through blogs and used a model of the spread of epidemics on networks~\cite{newman02} to characterize the spread of topics through the blogosphere.
\cite{Leskovec07blogs} defined an information cascade as a graph of hyperlinks between blog posts. A cascade starts with a cascade initiator, with other blog posts joining the cascade by linking to the initiator or other members of the cascade. Leskovec et al. found that the distribution of cascade sizes follows a power law.
%While \cite{Domingos01} developed algorithms to decide \emph{where} to initiate a cascade in a social network such that the cascade will be as large as possible, \cite{Leskovec07kdd} presented a method to decide which node to watch in a blog network in order to quickly detect large cascades.
In these studies, the networks were derived from the observed links between blog posts, i.e., from the diffusion of information. In our study, on the contrary, they were extracted from the sites independently of data about the diffusion of information. \cite{bakshy09} traced the spread of influence in a multi-player online game and found that similar to our findings with social news, influence spreads easily on social networks in virtual worlds. This provides an independent confirmation of the importance of social networks in the dynamics of information flow.

\section{Conclusion}
\label{sec:conclusion}
We conducted an empirical analysis of user activity on Digg and Twitter. Though the two sites are vastly different in their functionality and user interface,
%have different goals and motivations for existence and vastly different user interfaces,
they are used in strikingly similar ways to spread information. First, on both sites users actively create social networks by designating as friends others whose activities they want to follow. Second, users employ these networks to discover and spread information, including news stories. The mechanism for the spread of information is the same on both sites, namely, users watch their friends' activities --- what they tweet or vote for --- and by their own tweeting and voting actions they make this information visible to their own fans or followers. In spite of the similarities, there are quantitative differences in the structure and function of social networks on Digg and Twitter. Digg networks are dense and highly interconnected. A story posted on Digg initially spreads quickly through the network, with users who are following the submitter also likely to follow other voters. After the story is promoted to Digg's front page, however, it is exposed to a large number of unconnected users. The spread of the story on the network slows significantly, though the story may still generate a large response from Digg audience.
The Twitter social network is less dense than Digg's, and stories spread through the network slower than Digg stories do initially, but they continue spreading at this rate as the story ages and generally penetrate the network farther than Digg stories.

% model of social dynamics, quality
Understanding characteristics of user activity and the effect social networks have on it will help us make better use of social media and peer production systems. Currently these systems blindly aggregate activities of all users in order to identify high quality contributions. However, since popularity and quality are rarely linked~\cite{Salganik06}, this method is likely to highlight popular, though trivial, contributions. Separating in-network and out-of-network user activity, however, will lead to a better understanding of social dynamics of peer production systems~\cite{Hogg09icwsm,HoggSzabo09icwsm,Lerman10www}, which will allow us to better separate high quality contributions from noise~\cite{Hogg10icwsm,CraneSornette08,Lerman08wosn}.

\remove{

\subsection{Prediction}
Does the signature of a story's spread through the social network indicate how popular the story will become?

Plot number of fan votes in the first 10 votes total number of votes, averaged over all stories (see Lerman \& Galstyan 2008)
}

\subsection*{Acknowledgments}
%Prashant Khanduri, Tad Hogg
We are grateful to Tad Hogg for valuable insights into data analysis and to Prashant Khanduri for initial analysis of Digg data. This material is based upon work supported by the National Science Foundation under Grant No. 0915678.

%\bibliography{references}
%\bibliographystyle{aaai}

\end{document}